\documentclass[a4paper, 12pt]{article}
\usepackage{amssymb}

\usepackage{amsmath}

\usepackage{graphicx}
\usepackage{tikz}
\usepackage{scalerel}
\usepackage{pict2e}
\usepackage{tkz-euclide}
\usepackage{fullpage}
\usetikzlibrary{calc}
\usepackage{tikz}
\usetikzlibrary{arrows,backgrounds}
\usetikzlibrary{shadows}
\usetikzlibrary{external}

\usepackage{bbm}
\usepackage{pgf}
\usepackage{natbib}
\usepackage{hyperref}
\hypersetup{colorlinks,
  citecolor=red,
  linkcolor=red,
  urlcolor=red}
\usepackage{authblk}
\usepackage{blindtext}

\usetikzlibrary{patterns}
 
\graphicspath{{desktop/Networks model for diversity}}
\title{Random Attention and Unobserved Reference Alternatives\footnote{I thank Prof. Debasis Mishra and Prof. Arunava Sen for their valuable guidance.
This paper greatly benefited from suggestions by Abhinash Borah, Bhavook Bhardwaj, Siddharth Chatterjee, Dinko Dimitrov, Sean Horan, Ojasvi Khare, Matthew Kovach, Marco Mariotti, Saptarshi Mukherjee, Shashank Shekhar and Vilok Taori as well as from the helpful comments of conference participants at ISI Delhi, Delhi School of Economics, AMES-CSW 2024, and IGIDR CoRe 2024.}}
\author{Varun Bansal\footnote{Indian Statistical Institute, Delhi, India. Email: bansalvarun1999@gmail.com}}

\newtheorem{theorem}{Theorem}

\newtheorem{definition}[theorem]{Definition}

\newenvironment{proof}[1][Proof]{\noindent\textbf{#1.} }{\ \rule{0.5em}{0.5em}}
\newtheorem{innercustomgeneric}{\customgenericname}
\providecommand{\customgenericname}{}
\newcommand{\newcustomtheorem}[2]{%
  \newenvironment{#1}[1]
  {%
   \renewcommand\customgenericname{#2}%
   \renewcommand\theinnercustomgeneric{##1}%
   \innercustomgeneric
  }
  {\endinnercustomgeneric}
}

\newcustomtheorem{customthm}{Theorem}
\newcustomtheorem{customlemma}{Lemma}
\newcustomtheorem{customex}{Example}

\begin{document}

\maketitle
\begin{abstract}
In this paper, I develop and characterize two models of random attention that differ from each other with respect to the menu-dependence of the unobserved reference alternatives. In both models, the decision-maker pays attention to subsets of the available set of alternatives randomly with the reference alternatives being always paid attention to. Under menu-dependence, partial identification of both the reference alternatives and the underlying preferences is provided. For the case of multiple menu-independent references, I provide a complete identification of the references and a coarse identification of the underlying preferences. A complete identification of the latter is provided when the independent random attention function is considered.
\end{abstract}
\section{Introduction}
Decision-making is often constrained by limited attention, whether due to cognitive processing limits caused by the availability of too many alternatives or to a shortage of time. In such situations, decision-makers (DM) often resort to heuristics that can help them decide relatively quickly with less mental exercise. If the DM has to make similar choices multiple times with the same underlying primitives, the heuristic that decides how to allocate attention can lead to different choices. Stochastic choice functions represent such choices, with the main research task then being to ascertain the likely underlying heuristic and the primitives that drive the decisions of the DM. \cite{Cattaneo} introduced the Random Attention Model (RAM) framework to deal with the analysis of stochastic choice functions with random choice due to limited attention.\\\\
A substantial body of literature has explored how an observable reference affects the choices of a DM. In particular, \cite{Kovach23} uses the framework of RAM to look at how a reference alternative, or a status quo in their case, can affect choices if it directs attention. These authors were the first to deal with observable reference alternatives' impact on attention in a stochastic choice setting. They provide a characterization of a reference-dependent RAM, as well as special cases where the attention function is further restricted.\\\\
In certain cases, it is logical to assume that the reference is observable. For instance, for a DM in the market for a new car, her current car can be a reference. However, in many situations, it is not realistic to assume the knowledge of the DM's reference to anyone else but the DM themselves. Further, to analyze a choice problem with observable references, we have to deal with a family of stochastic choice functions, as done by \cite{Kovach23}. The inputs for the stochastic choice functions in their model were pairs of subsets\footnote{Hereafter referred to as menus.} and references- 
one family of stochastic choice function for each alternative as the reference. This is quite a significant data requirement. If we restrict the references to be unknown, we deal with just a single stochastic choice function over menus. \\\\
It is also not obvious that a reference alternative must be unique. In one of the models presented in this paper, I allow for multiple reference alternatives; clearly, one particular such alternative could be the status quo. In this spirit, I define an alternative to be a reference alternative if and only if it is \textit{attention privileged}, i.e. it is always paid attention to. The first model I look at has for each menu, an unobservable reference alternative that directs attention, and each menu can potentially have any alternative as the reference: the general model makes no restriction on how the reference changes with changing menus. \\\\
The second model allows for multiple reference alternatives. In particular, it defines multiple menu-independent references, i.e., there is a set of alternatives that are always attention privileged whenever offered. I provide four examples below that relate to the assumptions that I make in this paper.
\begin{itemize}
    \item \textbf{Example 1:} \textbf{University Choice:} High School students making a choice of university to continue their higher education in the UK filter out the large number of choices into smaller consideration sets. The universities are divided into \textit{old} and \textit{new}. This exogenous distinction impacts what universities are considered by students. For instance, \cite{Dawes} found that minority ethnic groups had a greater proportion of old universities in their consideration set, possibly due to reputation concerns of these ethnic groups. The college choice of high school students in the US is also a rather elaborate process. \cite{Rosen} showed that what major is offered as well as location, are crucial factors in shortlisting colleges. Parental and guidance counselor advice also has a significant impact on the decision-making process. It is conceivable that colleges that are close by act as references. Given the choice of colleges, the reference then changes depending upon the menu.
    \cite{Dawes} observed a similar phenomenon in the UK data as well. If the guidance of a counselor or a parent impacts which colleges act as references, it is not possible to observe these references. 

    \item \textbf{Example 2:} \textbf{Apparel Shopping:} Brand name is often crucial in drawing attention to particular items when it comes to shopping for apparel. The Air Jordan always catches the eye due to its association with Michael Jordan. However, though brand awareness might be enough to pull attention, it is not enough to provide a positive brand association as found by \cite{Dew}. Thus, attention and preference seem to be independent attributes. If attention is brand-dependent, it would make sense that attention privilege is menu-independent: Air Jordans are always paid attention to when available.
    

  \item \textbf{Example 3: Political Choice:} A stochastic choice framework can also be used to model a population of voters with similar preferences but with access to different options. A nationally strong party will always be considered as a reference; however, a weaker party in one district might be a stronger party in another and be considered there, leading to different choices even when voters exhibit homogenous preferences. So, some alternatives might be references whenever available, while some alternatives' being references are context dependent. This observation has been supported by \cite{Wilson}'s analysis of the Mexico national elections in 2000. The biggest party in Mexico, the Partido Revolucionario Institucional (PRI) has a strong presence in all the states, which makes it the reference, while the two opposition parties have varied presence in different states.
   

\item \textbf{Example 4:} Suppose past choices of the DM influence their future choices, i.e. they act as a reference. An outside observer can have information about the current feasible sets and the choices from those, but not the choices in the past. In this case, (multiple) past choices act as references and thus, they are unobservable for everyone but the DM. If the reference depends on recommendations from advisors (college choice) or past choices in elections (political choice), it is possible that the researcher is not able to observe the reference.\footnote{I thank Marco Mariotti for providing such an interpretation of the model.} 
\end{itemize}
In this paper, I look at two generalizations of the random attention model: one where each menu is endowed with an unobservable reference which is attention privileged (\textbf{GRAM-UR)}; and another with unobservable menu-independent, possibly multiple reference alternatives that are attention-privileged \textbf{(RAM-UR)}. I provide a characterization of the GRAM-UR and the RAM-UR by using testable axioms and identify the corresponding primitives (i.e., the set of reference alternatives and the underlying preferences). In the GRAM-UR, multiple representations are possible and references change depending upon the representation; however, one can uniquely identify the set of references for the RAM-UR, but the underlying preferences are only coarsely identified. I then restrict the attention function to be of the independent random attention (IRA) form (\cite{MM14}) and show that under such a restriction, the preferences can be uniquely identified. The existence of menus with no references is crucial in making the preference identification in this case. This is possible because of the menu independence of the references. \cite{Kovach23} allowed for the references to be part of the primitive, which precludes them from considering menus without references\footnote{In the GRAM-UR, menus without references are not possible, but they are possible in the RAM-UR and the RAM-UR-IRA.}. It is also notable that rational choice is a boundary case of the RAM-UR and the RAM-UR-IRA, while the IRA model is itself a boundary case of the RAM-UR-IRA\footnote{Boundary cases can be of two types the RAM-UR or RAM-UR-IRA: When the set of references is the set of all alternatives, or when the set of references is empty.}. Rational choice is also a special case of the GRAM-UR.\\\\
The rest of the paper is organized as follows. In Section 2, I introduce the generalized RAM-UR (GRAM-UR) model, the axiomatic characterization along with a discussion on the identification results. Section 3 introduces and characterizes the RAM-UR and discusses the identification results, compares them to the GRAM-UR and past models of limited attention.
The introduction and characterization of the RAM-UR-IRA are then provided in Section 4, where I also discuss the nestedness properties of the corresponding random attention models. Section 5 relates my work to the existing literature. All proofs are relegated to Appendix 1.
\section{Generalized Random Attention Model with Unobserved References}
Let $X$ be the finite set of all possible alternatives. Denote the set of all subsets of $X$ by $\mathcal{X}$, with a typical element denoted by $A$.
\begin{definition}
 A stochastic choice function is a mapping $p:X\times \mathcal{X}\setminus\phi\rightarrow [0,1]$ where for each $A\in\mathcal{X}\setminus\phi$ the following three conditions hold:
\begin{enumerate}
    \item $p(a,A)\geq 0$ for all $a\in X$
    \item $p(a,A)=0$ for all $a\in X\setminus A$
    \item $\sum_{a\in A}p(a,A)\leq1$
\end{enumerate}   
\end{definition}
The third part of the above definition allows for the sum of choice probabilities of alternatives from a menu not to sum up to one. I denote by $a^{*}$ the default alternative and interpret it as abstention, failing to make a choice or delaying choice. The idea of default is common in the literature on attention, especially the literature originating from \cite{MM14}. The assumption is that DMs prefer to get something over abstention\footnote{Equivalently, the default is only chosen if none of the alternatives is paid attention to.}. For each $A\in\mathcal{X}\setminus\phi$, define $p(a^{*},A)=1-\sum_{a\in A}p(a,A)$ to be probability of abstention from the menu $A$. I will also denote for any $A\in\mathcal{X}\setminus\phi$, $A^{*}=A\cup\{a^{*}\}$. The inclusion of the default in a choice model is not ideal, however, since the probability of default is just the residual probability in case the observable choice probabilities do not sum up to one, analysis is possible if we restrict the model to have probabilities sum up to one; we do that in GRAM-UR.\\\\
The formal definition of a random attention model uses the notion of an attention function.
\begin{definition}
An attention function is a mapping $\mu:\mathcal{X}\times\mathcal{X}\rightarrow[0,1]$ such that:
\begin{enumerate}
    \item $\sum_{D\subseteq A}\mu(D,A)=1$
    \item $\mu(D,A)>0$ if and only if $D\subseteq A$
\end{enumerate}
\end{definition}
Here $\mu(D,A)$ can be interpreted as the probability that $D$ is the consideration set when the menu offered is $A$. It is natural that if $D$ is not a subset of $A$, then the function should take a value of zero. The second point also stresses the fact that each subset of $A$ can be a consideration set with some positive probability (\textit{full support assumption}). In contrast to the above definition, \cite{Cattaneo} impose a monotonicity condition on the attention function and do not impose the full support assumption. \cite{Kovach23} imposed the full support assumption. It will be crucial in our model as well to identify the set of references. A modified full support assumption in line with the idea that references are exactly those alternatives that are attention-privileged is quite natural. I also relax the monotonicity condition in my models. 
A random attention model (RAM) is then defined as follows.
\begin{definition}
A stochastic choice function $p$ is a Random Attention Model (RAM) if there exists an attention function $\mu: \mathcal{X}\times\mathcal{X}\rightarrow[0,1]$ and a strict total order $\succ$\footnote{A strict total order is an asymmetric, transitive and complete binary relation defined over $X$.} over $X$ such that for all $A\in\mathcal{X}\setminus\phi$ and $x\in A$:
$$p(x,A)=\sum_{D\in\{D\in\mathcal{X}:x\in D, x\succ y, \forall y\in D\setminus\{x\}\}}\mu(D,A)$$
\end{definition}
In other words, a random attention model is a stochastic choice function where for each menu, each possible subset of a menu (consideration sets) is considered/paid attention to with some positive probability. The probability that an alternative is chosen from a menu is the sum of the attention probabilities of consideration sets where that alternative is the most preferred with respect to the underlying preferences.\\\\
Since the empty set is also a possible consideration set for each menu, the probability of choosing the default is just the probability of the empty set being the consideration set.
Note that the RAM as defined above without the full support assumption has no empirical content since we can just attribute the probability of each alternative in a menu to the singleton menu containing that alternative and the probability of abstention to the attention probability of the empty set.\\\\
In the generalized random attention model with unobserved references, each menu comes endowed with a reference alternative, which is attention-privileged. The attention function then takes the following form:
\begin{definition}
Let $r_{A}\in A\subseteq X$ denote the reference alternative for the menu $A$. A generalized reference-dependent attention function is a mapping $\mu_r:\mathcal{X}\times\mathcal{X}\rightarrow[0,1]$ such that:
\begin{enumerate}
    \item $\sum_{D\subseteq A}\mu_r(D,A)=1$.
    \item $\mu_r(D,A)>0$ if and only if $\{r_A\}\subseteq D\subseteq A$\footnote{This is the modified full support assumption, every subset which maintains the attention privilege
of the reference is paid attention to with positive probability.}.
\end{enumerate}
\end{definition}
The idea here is similar to attention privilege as described by \cite{Kovach23}, i.e. a reference is always considered. In fact, that is how a reference is defined in this setting. An alternative is a reference if and only if it is attention privileged. This definition can be extended to allow for multiple references, which we do in the model in section 3. Note that if there are no references in any menu, this definition is the same as a standard attention function as defined in Definition 2.\\\\
The modified full support assumption says that inherently, there is no bias towards any set when it comes to attention, apart from the bias to always consider the reference. The reference being menu-dependent means that the reference itself could be context-dependent: for instance, the biggest item could be the reference amongst any set of alternatives, which means that the reference keeps changing based on the set. However, I do not impose any consistency requirement on the reference across menus in the general model.\\\\
The fundamental analytical task with respect to the presented model is to identify the reference alternative and the underlying preferences. In most canonical papers, it is assumed that the references are observable, which precludes the need to identify the set of references. In the RAM model introduced by \cite{Cattaneo}, the monotonicity condition is crucial in making a partial identification of the underlying preferences. The (partial) identification of the preferences in the current model is aided by the full support assumption and the presence of attention-privileged alternatives.
\begin{definition}
A stochastic choice function $p$ is a \textbf{generalized random attention model with unobserved references (GRAM-UR)} if for all menus there exists a reference alternative $(r_A)_{A\subseteq X}$, a strict total order $\succ$ over $X$ and a generalized reference-dependent attention function $\mu_r:\mathcal{X}\times \mathcal{X}\rightarrow[0,1]$ such that for all $A\in \mathcal{X}\setminus\phi$ and $x\in A$:
$$p(x,A)=\sum_{D\in\{D\in\mathcal{X}:x\succ y,\forall y\in D\setminus\{x\}\}}\mu_r(D,A)$$
\end{definition}
The above definition is similar to the definition of RAM, apart from the fact that it allows for a generalized reference-dependent attention function. \\\\
A feature of the GRAM-UR is that the default alternative is never chosen with a positive probability in any menu. This is because the reference acts as the de facto default, i.e. whenever one of the other alternatives is not chosen, the DM sticks with the default. This property of the reference is similar to its function as a status quo, as modeled by several papers like \cite{MO05} and \cite{MO14}. \\\\
The partial identification of underlying preferences relies on the fact that any alternative that is not chosen with positive probability in a menu is not preferred to the attention-privileged reference, and any alternative that is chosen with a positive probability beats the reference. In the next section, I describe an axiomatic characterization of the GRAM-UR using testable axioms that ensure consistency of the preferences elicited in the way described above\footnote{The elicited preferences are inconsistent if they violate asymmetry or transitivity.}.  
\subsection{Characterization}
In this section, I characterize the GRAM-UR using two testable axioms.\\\\
\textbf{Acyclicity of No Choice (ANC):} Consider a sequence of alternatives $x_1,x_2,...,x_k\in X$, $k\geq 2$ such that there exist $A_1,A_2,...,A_{k-1}\in \mathcal{X}$ with $p(x_i,A_i)>0$ and $p(x_{i+1},A_i)=0$ for all $i\in\{1,2,...,k-1\}$, then there does not exist $A_k\in\mathcal{X}$ such that $p(x_k,A_k)>0$ and $p(x_1,A_k)=0$.\\\\
This is an axiom similar to one imposed by \cite{Cattaneo} to characterize the RAM. It disciplines the way a probability-zero choice is observed. In the GRAM-UR, the full support condition implies that zero probability choice for an alternative signifies that the reference is preferred to the alternative. In the same menu, if another alternative is chosen with a positive probability, then the reference cannot be preferred over it. ANC ensures we do not observe a cycle of such probabilities.\\\\
\textbf{No Default (ND):} For each $A\in\mathcal{X}$, $\sum_{x\in A}p(x,A)=1$.\\\\
The ND condition is usually assumed in the primitives of choice models. I have introduced it as an axiom instead to allow the GRAM-UR to lie in the same framework as other attention models like that of \cite{MM14} and the other models I introduce later in the paper.
\begin{customthm}{1}
A stochastic choice function is representable by a generalized random attention model with unobserved references if and only if it satisfies ANC and ND.    
\end{customthm}

Note that if a stochastic choice function has a GRAM-UR representation, such a representation is not unique. The proof in Appendix 1 shows that there are possible extensions of the partially identified preferences (\cite{Szpilrajn}) each of which provides a representation. However, when an extension is fixed, the reference for each menu is uniquely identified. Given that we are working with unobservable references and significant restrictions to the primitives as compared to \cite{Kovach23}, it is natural that the preference identification result is weaker. However, it is not weaker than the identification results of the RAM by \cite{Cattaneo}. 
\subsection{Multiple References Interpretation}
A possible interpretation of the GRAM-UR model is for each menu to have multiple reference alternatives that are attention-privileged. However, in such a case, the most preferred reference alternative in any menu is the de facto default, which prevents any of the other references to be chosen with positive probability. This means that one possible representation for a GRAM-UR could be where all alternatives that are chosen with zero probability and the lowest preferred alternative (according to the extension) are all references. This representation of the GRAM-UR incorporates the idea of multiple reference alternatives.
The attention function associated with the multiple reference interpretation takes interesting forms. For instance, the function that maps each menu to the set of references it contains could satisfy the attention filter by \cite{Masatlioglu12}, or the competition filter by \cite{Lleras}.
For the example below,  partial identification of preferences as well as the references is possible.

\begin{customex}{5}
Consider the set of alternatives $X=\{a,b,c\}$ and the following observed choice probabilities:
$$p(b,\{a,b\})=1/2\hspace{5mm}p(c,\{a,c\})=1/2 \hspace{5mm}p(a,\{a,b\})=p(a,\{a,c\})=1/2$$
$$p(c,\{a,b,c\})=p(b,\{a,b,c\})=1/2$$
$$p(b,\{b,c\})=1\hspace{5mm}$$
Here, the choice from $\{b,c\}$ gives us that $b$ is the reference in that menu and that $b$ is preferred to $c$. Since $a$ is not chosen with a positive probability in $\{a,b,c\}$, we have that $c$ is preferred to $a$. Here, in fact, we can completely and uniquely identify the underlying preferences: $b\succ c\succ a$ with the reference in each set being: $\{a,b\}\rightarrow a, \{a,c\}\rightarrow a,\{b,c\}\rightarrow b,\{a,b,c\}\rightarrow c$.\\\\
The multiple references interpretation allows that in $\{b,c\}$ the set of references is $\{b,c\}$ and in $\{a,b,c\}$ the set of references is $\{a,c\}$.
\end{customex}

In the next section, I look at a model focused on multiple references as the primitive with an additional restriction: the set of references is menu-independent. I provide a characterization of the model and show that in such a model, the set of references can be uniquely identified.
\section{Random Attention Model with Unobserved References: Fixed Set of References}
In this section, I introduce the reference-dependent random attention model with the restriction that the set of references is menu-independent. Here, there exists a set of alternatives such that whenever an alternative from that set is available in a menu, it is attention privileged.\\\\
I start by first introducing the notion of a reference-dependent attention function. 
\begin{definition}
A reference-dependent attention function with respect to a set $E\subseteq X$ of references is a mapping $\mu_{E}:\mathcal{X}\times\mathcal{X}\rightarrow [0,1]$ such that:
\begin{enumerate}
    \item $\sum_{D\subseteq A}\mu_{E}(D,A)=1$
    \item $\mu_{E}(D,A)>0$ if and only if $A\cap E\subseteq D\subseteq A$\footnote{This is another version of the modified full support assumption, every subset which maintains the attention privilege of the references is paid attention to with positive probability.}
\end{enumerate}
\end{definition}
In this model, multiple reference alternatives or no reference alternatives can be present in any menu, both of which are not possible in the model presented by \cite{Kovach23} or in GRAM-UR. Note that the default is only chosen in case none of the alternatives is paid attention to, i.e. if $\mu_{E}(\phi, A)>0$. This can only happen if $A\cap E=\phi$ and allows us to interpret the default as the de facto reference.
\begin{definition}
A stochastic choice function $p$ is a \textbf{random attention model with unobserved references (RAM-UR)} if there exists a set $E\subseteq X$, a strict total order $\succ$ over $X$ and a reference-dependent attention function $\mu_{E}:\mathcal{X}\times\mathcal{X}\rightarrow[0,1]$ such that for all $A\in\mathcal{X}\setminus\phi$ and $x\in A$:
$$p(x,A)=\sum_{D\in\{D\in\mathcal{X}:x\succ y, \forall y\in D\setminus\{x\}\}}\mu_{E}(D,A)$$
\end{definition}
\subsection{Characterization}
In the present section I characterize the RAM-UR by using three testable axioms. These axioms restrict the way in which the stochastic choice function operates across different menus.\\\\
\textbf{Existence of a Dominating Alternative (EDA):} For each $A\in\mathcal{X}\setminus\phi$: $p(x,A)=0$ for some $x\in A^{*}$ implies $p(y,\{x,y\})=1$ for some $y\in A\setminus\{x\}$.\\\\
A DM with limited attention fails to choose an alternative in two cases. First, if that alternative is never considered. Second, if it is considered, and is beaten by another alternative. EDA says that, the only way an alternative is never chosen is if another alternative always dominates it and is always considered in the binary choice between the two. This means that it is not possible for an alternative to never be chosen just because it is never considered, making each alternative to be non-trivial in the problem. The axiom also says that the set of possible consideration sets is rich enough that the only way an alternative is never chosen is if another dominating alternative is always considered, i.e. it is attention privileged. In this way, EDA captures the essence of the modified full support assumption.\\\\
%
\textbf{Certainty WARP (C-WARP):} For all $x\in X^{*}$ and $y\in X$ with $\{x,y\}\subseteq A^{*}\subseteq X^{*}$: $p(y,A)=1$ implies $p(x,A')=0$ for all $A'\subseteq X$ with $\{x,y\}\subseteq A'^{*}\subseteq X^{*}$.\\\\
Certainty WARP captures the idea that if an alternative is revealed to be dominated by an attention-privileged alternative, then it will never be chosen with a positive probability whenever that alternative is present. I call it C-WARP because it captures such revealed preference only when an alternative is chosen with certainty from a menu.\\\\
\textbf{Expansion (EXP):} For all $A,B\subseteq X$: $p(x,A)=p(y,B)=1$ for some $x\in A$ and $y\in B$ implies either $p(x,A\cup B)=1$ or $p(y,A\cup B)=1$.\\\\
Expansion captures the idea that as long as we keep observing certain choices in menus, there exists an attention-privileged alternative that dominates each alternative in the union of these menus. If we look at two attention-privileged alternatives, they will both be always considered, so the preference will determine which will be chosen, but it will be chosen with certainty in the union. It is in some ways similar to the expansion condition of \cite{MM07} in that it deals with the union of two menus. However, it works with any two menus with certain choices. It is silent on menus where certain choice is not observed.\\\\
The three axioms are independent, and they impose restrictions on stochastic functions only when deterministic choice or non-choice is observed. This is because the model is very flexible on the structure of the attention function, once the full support assumption is satisfied. To further restrict the stochastic function some regularity condition similar to \cite{Cattaneo} must be imposed. However, such an imposition is not necessary to get the first set of results described below.\\\\
The result below shows that the above three axioms do allow a stochastic choice function to be represented by a RAM-UR model. However, the representation might not be unique.
\begin{customthm}{2}
A stochastic choice function is representable by a random attention model with unobservable references if and only if it satisfies EDA, C-WARP and EXP.
\end{customthm}
It is worth noting that one can obtain multiple representations of a stochastic choice function by slightly modifying the constructed attention function and the extension of the partially identified preferences. However, in each representation, the set of reference alternatives will be uniquely identified, i.e. as compared to the GRAM-UR, the identification of references is better. In the GRAM-UR, the reference identified for each menu was dependent upon the extension of the underlying preference relation. In contrast, here it is independent of it.\\\\
Even with limited data about choices from menus, in most cases we can identify the set of references uniquely. For instance, if there is no data available on singleton sets but we have the choice probabilities from all two and three-element menus, identification will be mostly possible. A rare exception is the case where the second least preferred alternative is not a reference but the least preferred alternative is. In such a case, we will not be able to ascertain which of the two will be the least preferred (and hence, a reference) and which is the second least preferred. However, we will be able to identify all other reference alternatives uniquely.\\\\
Our identification result is that we are only able to identify the preferences coarsely, i.e. we can identify the relative ranking of an alternative (whether a reference or not) with respect to all reference alternatives as well as the transitive closure of the partial order constructed through this relative ranking. However, for any two (non-reference) alternatives that have the same relative rankings with respect to all the reference alternatives, we are not able to identify the preference and thus, one can represent the choice function by taking any extension (\cite{Szpilrajn}) of the revealed strict partial order\footnote{A strict partial order is a transitive, irreflexive and asymmetric binary relation defined over $X$.}. The coarse identification is the result of the restrictions on both the observability of the references and on the amount of data available.\\\\
Even though the axioms impose restrictions only when deterministic choice or non-choice is observed, it is not necessary to observe certain choice to identify the set of references. It is also not necessary to observe all menus to make such an identification. The following example illustrates this point:
\begin{customex}{6}
Consider the set of alternatives $X=\{a,b,c,d\}$ and the following observed choice probabilities: 
$$p(a,\{a,b,c\})=0.4\hspace{5mm}p(b,\{a,b,c\})=0.3 \hspace{5mm}p(c,\{a,b,c\})=0.3$$
$$p(a,\{a,c,d\})=0.6\hspace{5mm}p(c,\{a,c,d\})=0.4 \hspace{5mm}p(d,\{a,c,d\})=0$$
$$p(a,\{a,b,d\})=0.4\hspace{5mm}p(b,\{a,b,d\})=0.3 \hspace{5mm}p(d,\{a,b,d\})=0.2$$
$$p(b,\{b,c,d\})=0.5\hspace{5mm}p(c,\{b,c,d\})=0.5 \hspace{5mm}p(d,\{b,c,d\})=0$$
Here we observe data only from menus of size three, and we do not observe any deterministic choice.
\end{customex}
Note that in the above example, we can identify alternative $c$ as the only reference, as well as the partial order without the need for the complete set of menus. In fact, in any setting with a large number of alternatives and relatively few references, we can make full identification with very few data points, and we will not need to observe a lot of deterministic choices in those menus.\\\\
\cite{Kovach23} were able to identify the complete underlying preferences from the general model with observed references. The RAM-UR works with stochastic choice functions and the references are unobservable. Given the significant restrictions to the primitives in this model,
it is natural that the preference identification result is weaker. 
However, even in the restricted model, quite a significant part of the preferences are identified. Even among non-reference alternatives, with different relative ranking with respect to the references, unique identification of the preferences among them is achieved.\\\\
A natural way to provide a full identification of preferences is to impose some restrictions on the structure of the attention function. First, I will show that the monotonicity condition imposed by \cite{Cattaneo} does not lead to better identification results. 

\begin{definition}
A reference-dependent monotonic attention function with respect to a set $E\subseteq X$ of references is a mapping $\mu_{E}:\mathcal{X}\times\mathcal{X}\rightarrow[0,1]$ such that:
\begin{enumerate}
    \item $\sum_{D\subseteq A}\mu_{E}(D,A)=1$
    \item $\mu_{E}(D,A)>0$ if and only if $A\cap E\subseteq D\subseteq A$
    \item For $S,T\in\mathcal{X}$, if $a\in S\setminus T$, then $\mu_{E}(T,S\setminus\{a\})\geq \mu_{E}(T,S)$
\end{enumerate}
\end{definition}
The example below illustrates that even if we restrict ourselves to monotonic attention functions, a unique identification of the preferences is not possible.
\begin{customex}{7}
Suppose we have three alternatives $X=\{a,b,c\}$ and the following observed choice probabilities:
$$p(b,\{a,b\})=1/2\hspace{5mm}p(c,\{a,c\})=1/2 \hspace{5mm}p(a,\{a,b\})=p(a,\{a,c\})=1/2$$
$$p(a,\{a,b,c\})=p(c\{a,b,c\})=1/4\hspace{5mm}p(b,\{a,b,c\})=1/2$$
$$p(a,\{a\})=1\hspace{5mm}p(b,\{b\})=p(c,\{c\})=1/2$$
$$p(b,\{b,c\})=1/4\hspace{5mm}p(c,\{b,c\})=1/2$$
We can rationalize these observed choice probabilities through two reference-dependent monotonic attention functions. In both cases $E=\{a\}$ and both $b$ and $c$ are revealed preferred to $a$.
\begin{itemize}
    \item $\mu^{1}_{E}$ is the function such that attention is divided among all eligible\footnote{Supersets of \{a\} in any menu containing $a$, and all subsets in any other menu.} sets in each menu with the preference being $b\succ c\succ a$.
    \item Construct $\mu^{2}_{E}$ as follows:
    $$\mu^{2}_{E}(\{a,b\},\{a,b,c\})=1/2\hspace{5mm}\mu^{2}_{E}(\{a\},\{a,b,c\})=1/4$$
    $$\mu^{2}_{E}(\{a,c\},\{a,b,c\})=\mu^{2}_{E}(\{a,b,c\},\{a,b,c\})=1/8$$ 
    $$\mu^{2}_{E}(\{a,b\},\{a,b\})=1/2\hspace{5mm}\mu^{2}_{E}(\{a,c\},\{a,c\})=1/2$$
    $$\mu^{2}_{E}(\{b\},\{b,c\})=1/8\hspace{5mm}\mu^{2}_{E}(\{b,c\},\{b,c\})=1/8\hspace{5mm}\mu^{2}_{E}(\{c\},\{b,c\})=1/2$$
    $$\mu^{2}_{E}(\{a\},\{a\})=1\hspace{5mm}\mu^{2}_{E}(\{b\},\{b\})=1/2\hspace{5mm}\mu^{2}_{E}(\{c\},\{c\})=1/2$$
    With the above attention function which is also monotonic and the preference $c\succ b\succ a$ we can also rationalize the observed choice probabilities.
\end{itemize}
\end{customex}
Monotonicity in itself is not enough to refine our identification results, however, further restrictions will be. In this direction, I characterize the RAM-UR-IRA by restricting the attention function to be of the Independent Random Attention (IRA) form as in \cite{MM14}. The RAM-UR-IRA then allows for a unique identification of the underlying preferences and the attention function.
\section{Random Attention Model with Unobserved References: Independent Random Attention}
In this section I assume that the attention function in the RAM-UR model is of an independent random type in the sense that there exists an attention parameter $\gamma: X\rightarrow(0,1]$ such that the attention function takes the form: $\mu_{IRA}(D,A)=\prod_{x\in D\subseteq A}\gamma(x)\prod_{y\in A\setminus D}(1-\gamma(y))$ for all $A\in \mathcal{X}\setminus\phi$. Here attention privilege manifests in the form of alternatives with the attention parameter to be one, which is the key difference between this form and the attention function of the IRA form introduced by \cite{MM14} (hereafter referred to as the IRA model). The IRA restricts the attention parameter to be strictly between zero and one while we do not impose such a restriction.
This makes the IRA model to be nested in the RAM-UR-IRA. \\\\
Here the set of attention-privileged alternatives $E$ is just the set containing all alternatives $x\in X$ such that $\gamma(x)=1$. Since the $\gamma$ function takes strictly positive values for all alternatives, the RAM-UR-IRA is a special case of the RAM-UR as the full support assumption will hold. 
\begin{definition}
A stochastic choice function $p$ is a \textbf{random attention model with unobservable references and independent random attention (RAM-UR-IRA)} if there exists a strict total order $\succ$ over $X$ and an attention parameter $\gamma: X\rightarrow(0,1]$ such that for all $A\in\mathcal{X}\setminus\phi$ and $x\in A$:
    $$p(x,A)=\gamma(x)\prod_{y\in A; y\succ x}(1-\gamma(y))$$    
\end{definition}
Imposing the above additional structure significantly changes how the probabilities can be defined in the RAM-UR-IRA model. Even though it is a special case of RAM-UR and nests\footnote{By nesting the IRA (rational) I mean that a special case of RAM-UR-IRA: the one where $E=\phi$ ($E=X$) will also be an IRA (rational) model.} both the IRA as well as classical rationality, it is mathematically a small departure from the IRA model: It allows $\gamma$ to take value one for some alternatives. In what follows, I provide a characterization of this model by using axioms similar to the characterizations of both the IRA and the RAM-UR and discuss the identification results. 
\subsection{Characterization}
I characterize the RAM-UR-IRA using the following five independent axioms: \\\\
\textbf{R-Asymmetry (R-ASYM):} For all $A\in\mathcal{X}\setminus\phi$ and $a,b\in A$ with $p(a,A)>0$, $p(b,A)>0$: $\frac{p(a,A\setminus\{b\})}{p(a,A)}\neq 1$ implies $\frac{p(b,A\setminus\{a\})}{p(b,A)}=1$.\\\\
This axiom is similar to the i-asymmetry axiom that characterizes the IRA (cf. \cite{MM14}). However, it restricts the denominator to be non-zero. Such a restriction is not needed in the IRA characterization, since all probabilities are assumed to be positive in that model. The axiom says that if removing an alternative $b$ changes the 
choice probability of another, say $a$ (given the positive denominator constraints), then removing $a$ would not change the choice probability of $b$. \\\\
\textbf{R-Independence (R-IND):} (1) For all $A, B\in\mathcal{X}\setminus\phi$ and $a,b\in A\cap B$ with $p(a,A)>0,p(a,B)>0$: $\frac{p(a,A\setminus\{b\})}{p(a,A)}=\frac{p(a,B\setminus\{b\})}{p(a,B)}$.
\\\\
(2) For all $A, B\in\mathcal{X}\setminus\phi$ and $b\in A\cap B$ with $p(a^{*},A)>0, p(a^{*},B)>0$: $\frac{p(a^*,A\setminus\{b\})}{p(a^*,A)}=\frac{p(a^*,B\setminus\{b\})}{p(a^*,B)}$.\\\\
This axiom is similar to the i-independence axiom in the characterization of the IRA (cf. \cite{MM14}), with the same constraints regarding non-zero probability when needed. It says that the ratio effect of removing an alternative on another alternative's choice probability (subject to the positive denominator constraints), is menu-independent, i.e. it is the same whatever be the menu from which it is removed.\\\\
\textbf{Non-Triviality (NT):} $p(x,\{x\})>0$ for all $x\in X$.\\\\
This axiom implies that each alternative has a value on its own in the sense that if this axiom is not satisfied and we have an alternative that is not chosen in a singleton menu with positive probability, then other axioms will ensure that this alternative will not be chosen with a positive probability in any menu. Note that EDA introduced in Section 2.2 implies NT.\\\\
\textbf{Existence of a Dominating Alternative* (EDA*):} For all $A\in\mathcal{X}\setminus\phi$: $p(a^{*},A)=0$, implies $p(a^{*},\{x\})=0$ for some $x\in A$.\\\\
This axiom is similar in spirit to the EDA axiom used in the characterization of RAM-UR. However, we only need a check for the default alternative as opposed to checking for all alternatives that are chosen with zero probability in a menu. Note again that EDA implies EDA*, which means that NT and EDA* together are weaker than EDA.\\\\
\textbf{Regularity (REG):} For all $A,B\in\mathcal{X}\setminus\phi$: $A\subset B$ implies $p(a,A)\geq p(a,B)$ for all $a\in A^{*}$.\\\\
This is the standard regularity axiom which states that an alternative's choice probability weakly reduces in larger menus. This property is in general not satisfied by the RAM-UR. If we consider the IRA model, the i-asymmetry and the i-independence axioms imply the regularity property. However, that is not the case with the other four axioms utilized to characterize the RAM-UR-IRA. Thus, the property is used as an axiom for the characterization. In the absence of non-triviality, REG implies that any trivial alternative\footnote{An alternative $x\in X$ here is called trivial if $p(x,\{x\})=0$.} will not be chosen with a positive probability in any menu.
\begin{customthm}{3}
A stochastic choice function is representable by a RAM-UR-IRA model if and only if it satisfies R-ASYM, R-IND, NT, EDA* and REG. Moreover, the representation is unique.    
\end{customthm}
Representation uniqueness means that the set of references, the underlying preferences, and the underlying attention parameters are uniquely identified. This uniqueness result is due to the additional structure imposed on the attention function and to the availability of data in menus with no reference points. \\\\
The proof of Theorem 3 essentially partitions the menus into menus with no references and such with at least one reference. Using the axioms above and the \cite{MM14} result on the set of menus with no reference, it identifies the underlying attention parameter and partial preferences among non reference alternatives (precisely the part of the preferences that were unidentified in the RAM-UR). Since the RAM-UR-IRA is a special case of the RAM-UR, the rest of the identification is easily done similar to the general model. The proof shows that the fitting of the identified preferences through the non-reference menus and the preferences identified through the menus with references does indeed give us a strict total order. Finally, using the first two axioms, we obtain the required functional form.\\\\
An advantage of a model like the GRAM-UR and RAM-UR is that it is in the realm of the stochastic choice function, which is why a nestedness analysis with some classical models can be undertaken. 
The GRAM-UR, the RAM-UR and the RAM-UR-IRA nest rational choice as a special case. The GRAM-UR, when for each menu, the most preferred alternative is the reference, is the same as rational choice. The RAM-UR and the RAM-UR-IRA are equivalent to rational choice when the set of references is the entire set of alternatives. The RAM-UR and the RAM-UR-IRA also nest the IRA, i.e. when the structure of the attention function is the IRA structure and the set of references is empty. Finally, the RAM-UR-IRA is also representable by a Random Utility Model (RUM) in a similar spirit to how the IRA can be. This is not true for the RAM-UR since it can violate regularity which is necessary for any RUM representation. A brief discussion on the RUM representation of the RAM-UR-IRA is provided in Appendix 2.\\\\
The no default restriction on the GRAM-UR prevents it from nesting the RAM-UR. However, with the multiple reference interpretation, restricting how references behave across menus can make these two versions quite close. I discuss the presence of default in attention models and how it might not be a bad assumption in the next section.
\section{Literature Overview}
The idea that limited attention affects behavior and choice is well accepted and widely documented in the literature. \cite{Simon} model of satisficing choice is one of the earliest models that can be interpreted as motivated by limited attention. In this model, the heuristic is that the DM stops at the first alternative which is ``good enough". A DM might utilize such a heuristic because of lack of time or attention. \cite{Lleras} modeled choice overload, which leads to worse outcomes in larger menus, the reason being that good options might fail to grab attention in larger sets due to an overload of available options. \cite{Masatlioglu12} introduce the concept of the attention filter which incorporates the idea that removing an unconsidered alternative does not change the consideration set in a deterministic choice setting where the DM is attention constrained. \cite{Dean17} is another paper that incorporates limited attention and also the effect of reference alternatives.\\\\
Models of stochastic choice ventured into modelling the variability observed in choices due to limited attention. \cite{MM14} is the seminal paper that filled this void. Since then there has been a substantial body of literature combining both limited attention and stochastic choice. A plethora of papers are closely related to \cite{MM14} and include, for instance, \cite{Brady}'s feasibility model as a generalization of the \cite{MM14}, though their motivation is not limited attention, but limited feasibility. \cite{Horan} looks at the IRA without observing the probability of default. \cite{Cattaneo} introduces the general idea of a random attention model. \cite{Kovach23} uses the idea of a RAM in the context of reference-dependent choice and models a family of attention functions, each indexed by the reference point, and looked at different special cases like the Constant Random Attention, Luce Random Attention and the Independent Random Attention. They discussed the nestedness properties among these special cases. Their model didn't lend itself to nestedness analysis with respect to the classical stochastic choice models because in their model, the primitives were menu-reference pairs. This placed it outside the realm of stochastic choice models.\\\\
The idea of reference-dependent choice has also been the focus of several papers. \cite{MO05} and \cite{MO14} are seminal papers in this area. These authors were one of the first to model status quo bias in choice, which is a widely observed behavior in experiments on individual choice\footnote{\cite{MO05} provides references for such experiments.}. \cite{Sagi06} imposes a no-regret condition on the reference dependent preferences
and \cite{Tapki07} allows for incomplete reference dependent preferences. Note that all mentioned works model the effect of the status quo on preferences. However, it is possible that the status quo might affect attention as well. \cite{Dean17} looks at status quo bias and limited attention together, allowing for both reference-dependent preferences and attention. \cite{Kovach23}, as discussed before, constructs a model where references affect only the attention. \cite{Bossert09} provides another model where reference points don't affect preferences. These authors model a deterministic choice function where the choice is consistent with not choosing a worse alternative than the reference. They study the existence of a linear order such that the choice from each menu-reference pair is at least as good as the reference point, though not necessarily the maximum\footnote{The best alternative according to the preference ordering.}. The GRAM-UR and the RAM-UR have this non-deteriorating property since references are attention privileged, and thus, in any menu, an alternative worse than the reference is never chosen.
\\\\
In the literature discussed above, a general theme is that the primitives include the reference point. It is natural to challenge the assumption of the observability of the reference point. \cite{OK15} and \cite{Tserenjigmid19} are two relevant papers here. However, in \cite{Tserenjigmid19} the references are not alternatives but a vector of attributes which depend upon the minimum of those attributes among the alternatives in any menu. It is thus, menu dependent as well. \cite{Kovach23} is the first paper to study reference dependence in a stochastic setting, which makes the current work the first to study unobservable references in this setup.\\\\
The literature on marketing and consumer research focused on consideration sets and the impact of memory and attention on whether certain brands enter consumers' consideration sets. \cite{Nedungadi} shows that whether an alternative is recalled while forming consideration sets is not necessarily correlated with how highly that alternative is evaluated by the consumer, i.e. attention might be independent of preferences. Advertising often plays a crucial role in increasing an alternative's possibility of consideration, by increasing its brand recall, or in the terminology of our paper, making it a reference. It is also possible that higher advertising, and thus, more attention, increases the probability of choice of another alternative. This is consistent with both the experiments of \cite{Huber} and \cite{Nedungadi}, and is in line with the model described in this paper. If a reference is added to a menu, the probability of choice of a different alternative can increase. The model described in this paper is relevant for more important decision problems like university choice for high school students (\cite{Dawes}, \cite{Rosen}) and political choice (\cite{Wilson}). \\\\
Finally, I would like to discuss the incorporation of a default alternative in the proposed framework. In models of limited attention, the default is used to model the possibility of choice avoidance, delaying choice, which is quite a well-established empirical observation (\cite{Dhar}). \cite{Horan} argued that it might not be possible to observe the probability of no choice from empirical choice data, and provided a characterization of the IRA by \cite{MM14} without the need for observing such probabilities. In a reference-dependent random utility model, the default is a de facto reference, and as such, any menu where another reference alternative is present, the default is not relevant. In the GRAM-UR, the presence of a reference in all menus makes it a standard stochastic choice problem without the possibility of choosing a default. A reference in any menu is the alternative against which any alternative that is considered is always evaluated. The separation of consideration and evaluation elements of the choice means that the default then becomes the reference for salience. The probability of choosing to abstain is the probability that none of the alternatives will be salient enough to draw attention. If there are other references present, those that are salient enough to always draw attention, the default is redundant.  
\section*{Appendix}
\subsection*{Appendix 1}
\textbf{Proof of Theorem 1}\\\\
\textit{Necessity:}\\\\
ANC: In the GRAM-UR, if an alternative, say $a$ chosen with a positive probability in a menu, while $b$ is not, then that reveals that $a$ is preferred to $b$. A cycle of the form that ANC prevents from existing corresponds with a cycle in this preference. Thus, ANC is necessary.\\\\
ND: Since each menu maps to a unique reference alternative in the RAM-UR, the empty set is considered with zero probability. In every other subset, some alternative is the maximal element and, thus, is chosen. So, the sum of choice probabilities of non-default alternatives sum up to one.\\\\
\textit{Sufficiency:}\\\\
Define the binary relation $\bar{\succ}$ as follows: $a\bar{\succ}b$ if there exists $A\in\mathcal{X}$ such that $a,b\in A$, $p(a,A)>0$ and $p(b,A)=0$. ANC with $k=2$ ensures that this binary relation is asymmetric, and ANC for arbitrary $k$ ensures that it is acyclic. Define by $\succ$ the transitive closure of $\bar{\succ}$. The above argument shows that $\succ$ is a strict partial order. Note that $\succ$ is the uniquely identified part of the underlying preferences.\\\\
Given that $\succ$ is a strict partial order we can extend it to a strict total order\footnote{A strict total order is an asymmetric, transitive and complete binary relation over $X$.} (\cite{Szpilrajn}). Such an extension need not be unique. Fix $\succ^*$ to be such an extension, that is, for all $x,y\in X$ we have that:

$$x\succ y\implies x\succ^*y$$

I define the set $D_x$ to be the largest subset of $X$ such that $x$ is the $\succ^*$-maximal element. Now we will define the revealed attention function for an arbitrary menu. Consider a menu $A\in\mathcal{X}$. Define by $r^*_{A}=argmin_{\succ^*}\{x\in A:p(x,A)>0\}$, i.e. the $\succ^*$-minimal element in $A$ such that it is chosen with a positive probability. We will call this the revealed reference. Note that for a given extension each menu maps to a unique reference.\\\\
Now the revealed attention function $\mu^*_{r}:\mathcal{X}\times\mathcal{X}$ is specified such that for each $x\in A$:

$$p(x,A)=\sum_{D\in \{D:\{r^*_{A}\}\subseteq D\subseteq D_x\cap A\}}\mu^*_{r}(D,A)$$

There are multiple ways to construct such an attention function. One way is as follows: For any arbitrary subset such that it does not contain $r^{*}_{A}$, the attention function takes the value zero. For any other subset, some alternative which has a positive choice probability is the maximal element, this is because it contains $r$ and it has a positive choice probability and any alternative which has a positive choice probability is preferred to it according to $\succ^*$. This is true because of the choice of $r^*_{A}$. We will determine the attention map for such a menu by dividing the choice probabilities of the $\succ^*$-maximal alternative among such menus. \\\\
Now we can complete the construction similar to that in \cite{Kovach23} for an arbitrary alternative $x$ which has a positive choice probability by the following procedure: Let $n(x,A):=|\{D:\{r^*_A\}\subseteq D\subseteq D_x\cap A\}|$. Then we specify $\mu^*_r(D,A)=\frac{p(x,A)}{n(x,A)}$ for all such $D\in\{D:\{r^*_A\}\subseteq D\subseteq D_x\cap A\}$. Note that this equal division of choice probabilities fulfills the modified full support assumption. Since, the stochastic choice function satisfies ND, the revealed attention function generated will be a generalized reference-dependent attention function according to Definition 4. The binary relation $\succ$ which is the transitive closure of $\bar{\succ}$ gives us the uniquely identified part of the underlying preferences.\\\\
\textbf{Proof of Theorem 2}\\\\
\textit{Necessity:}\\\\
EDA: Note that an alternative $x$ is never chosen from a menu if whenever it is paid attention to, a better alternative is paid attention to as well, which is only possible if a better alternative, say $y$ is attention privileged. In a binary menu including $x$ and $y$, $y$ will again be attention privileged so EDA will hold.\\\\
C-WARP: if $y$ is always chosen in a menu, then it is attention-privileged, and better than any other alternative because if it wasn't, the full support assumption would mean that the other alternative would be chosen with positive probability. So, C-WARP will be satisfied.\\\\
EXP: If the antecedent of the axiom holds, then both $y$ and $x$ are attention-privileged, and every other alternative in $A^*$ and $B^*$ is less preferred to one of the two, so will not be chosen with positive probability. If $x$ is preferred to $y$ then it will always be chosen, and vice versa. So, EXP holds.\\\\
%
%
\textit{Sufficiency:}\\\\
I take the set $\Bar{E}=\{x\in X|p(x,\{x\})=1\}$ to be the revealed set of reference alternatives.
If This set is empty EDA ensures that for any $A\in\mathcal{X}\setminus\phi$, $p(a,A)>0$ for all $a\in A^{*}$. In this case, we can construct an attention function with $E=\Bar{E}=\phi$ and an arbitrary preference. Note that with an empty set of references, the RAM-UR is just a RAM, so it suffices to show a RAM representation. For this, I take an arbitrary preference and construct an attention function as follows: For any alternative the probability that it is chosen is taken to be the sum of the value that the attention function takes on subsets where that alternative is the best. One way to do it is divide the probability of an arbitrary alternative $x$ in a menu $A$ to all sets in $\{D\subseteq A:x\succ y, \forall y\in D\setminus\{x\}\}$ equally ($\succ$ is the arbitrary preference). Note that since all alternatives (and default) are chosen with positive probability in all menus such a construction will give us a valid attention function. Given the constructed attention function, the observed choices are represented by a RAM with the attention function and the arbitrary preferences.
Such a construction will be one possible RAM representation. We can have multiple representations if we change how we assign the probabilities.
Thus, we have a representation when $\Bar{E}$ is empty. For the rest of the proof I assume that $\Bar{E}$ is non-empty.\\\\
In such a case, I define the binary relation $P$ over $X$ as follows:
\begin{enumerate}
    \item For $x\in \Bar{E}$ and $y\in X^{*}$ with $x\neq y$: $xPy$ iff $p(x,\{x,y\})=1$.
    \item For $x\notin \Bar{E}$ and $y\in \Bar{E}$: $xPy$ iff $p(x,\{x,y\})>0$.
    \item For $x,y\notin \Bar{E}$: $xPy$ iff there exists a $z\in \Bar{E}$ such that $xPz$ and $zPy$.
\end{enumerate}
\begin{customlemma}{1}
$P$ is a strict partial order over $X$.    
\end{customlemma}
\begin{proof}
To show the asymmetry of $P$, we need to show that it is antisymmetric and irreflexive. I take it case-wise:\\\\
If $x\in \Bar{E}$, then I haven't defined $xPx$. If $x\neq y$, then $xPy$ implies $p(x,\{x,y\})=1$. So, $yPx$ is not possible since $p(y,\{x,y\})=0$, and $yPx$ cannot be the case. So for this case, there is no violation of irreflexivity and antisymmetricity.\\\\
If $x\notin \Bar{E}$, then $xPx$ is only possible if there exists a $z\in \Bar{E}$ such that $xPz$ and $zPx$ but that is not possible as above, so $P$ is irreflexive. Now, if $y\neq x$ such that $y\in \Bar{E}$, then $xPy$ implies $p(x,\{x,y\})>0$, and thus, $yPx$ is not possible. Finally, if $y\neq x$ such that $y\notin \Bar{E}$, then $xPy$ implies the existence of $z\in \Bar{E}$ such that $p(x,\{x,z\})>0$ and $p(z,\{z,y\})=1$. If it were possible that there exists $w\in \Bar{E}$ such that $p(w,\{w,x\})=1$ and $p(y,\{w,y\})>0$, then by EXP we have either $p(w,\{w,x,y,z\})=1$ or $p(z,\{w,x,y,z\})=1$ in contradiction to C-WARP. So, $P$ is asymmetric.\\\\
Now I show the transitivity of $P$. I will again argue casewise, in each of the cases having $xPy$ and $yPz$ for $x,y,z\in X$:
\begin{itemize}
    \item \textbf{Case 1:} $x\in \Bar{E}$, $y\in \Bar{E}$ and $z\in X^{*}$.\\\\
     By the definition of $P$ we have $p(x,\{x,y\})=1$ and $p(y,\{y,z\})=1$. EXP and C-WARP give us $p(x,\{x,y,z\})=1$, applying C-WARP again gives us $p(x,\{x,z\})=1$, so we have $xPz$.
    \item \textbf{Case 2:} $x\in \Bar{E}$, $y\notin \Bar{E}$ and $z\in \Bar{E}$.\\\\
    We have $p(x,\{x,y\})=1$ and $p(y,\{y,z\})>0$. Note that $p(z,\{z\})=1$ since $z\in \Bar{E}$. So, by EXP, either $p(x,\{x,y,z\})=1$ or $p(z,\{x,y,z\})=1$, but the latter along with $p(y,\{y,z\})>0$ violates C-WARP, so we must have $p(x,\{x,y,z\})=1$ which by C-WARP gives us $p(x,\{x,z\})=1$ and thus, $xPz$.
    \item \textbf{Case 3:} $x\in \Bar{E}$, $y\notin \Bar{E}$ and $z\notin \Bar{E}$.\\\\
    We have $p(x,\{x,y\})=1$, as well as the existence of $w\in \Bar{E}$ such that $yPw$ and $wPz$. Note that $w\neq x$ since $P$ is asymmetric. Looking at the menu $\{x,w,z\}$ we have $p(x,\{x,w,z\})=1$ because by Case 2 we know that $xPw$ and by EXP we know that either $p(x,\{x,w,z\})=1$ or $p(w,\{x,w,z\})=1$ holds. By C-WARP, $xPz$ follows.
    \item \textbf{Case 4:} $x\notin \Bar{E}$, $y\in \Bar{E}$ and $z\in \Bar{E}$.\\\\
    We have $p(x,\{x,y\})>0$ and $p(y,\{y,z\})=1$. If $p(x,\{x,z\})=0$, then we have $p(z,\{x,z\})=1$ by EDA. By EXP, either $p(y,\{x,y,z\})=1$ or $p(z,\{x,y,z\})=1$ must be the case, so $p(x,\{x,y,z\})=0$. By C-WARP, it must be that $p(z,\{x,y,z\})=1$, and applying C-WARP again we get $p(z,\{y,z\})=1$ in contradiction to $yPz$. So, it must be that 
$p(x,\{x,z\})>0$ and thus, $xPz$ follows.
\item \textbf{Case 5:} $x\notin \Bar{E}$, $y\in \Bar{E}$ and $z\notin \Bar{E}$.\\\\
We have $xPy$, $yPz$ and $y\in \Bar{E}$, so by definition $xPz$ follows.
\item \textbf{Case 6:} $x\notin \Bar{E}$, $y\notin \Bar{E}$ and $z\in \Bar{E}$.\\\\
We have $xPy$ and thus, there exists $w\in \Bar{E}$ such that $xPw$ and $wPy$. By Case 2 we have $wPz$, and by Case 4 we have $xPz$.
\item \textbf{Case 7:} $x\notin \Bar{E}$, $y\notin \Bar{E}$ and $z\notin \Bar{E}$.\\\\
We have $xPy$ and thus, there exists $w\in \Bar{E}$ such that $xPw$ and $wPy$. So, by Case 3 we have $wPz$. By Case 5 $xPz$ follows.
\end{itemize}
In other words, the binary relation $P$ defined above is transitive. Note that P is complete when it comes to comparisons between an alternative in $\Bar{E}$ and any other alternative. The only comparisons it could be silent on are between alternatives not in $\Bar{E}$. 
\end{proof}\\\\
Given that $P$ is a strict partial order we can extend it to a strict total order (\cite{Szpilrajn}). Note that such an extension need not be unique.
Fix $\succ^{*}$ to be such an extension, that is, for all $x,y\in X$ we have that:
$$xPy\implies x\succ^{*}y$$
Next, I define the set $D_{x}$ to be the largest subset of $X$ such that $x$ is the $\succ^{*}$-maximal element.\\\\
Now I will specify the revealed attention function $\mu^{*}_{\Bar{E}}:\mathcal{X}\times\mathcal{X}$ such that:
$$p(x,A)=\sum_{D\in\{D:[x\cup(A\cap \Bar{E})]\subseteq D\subseteq D_{x}\cap A\}}\mu^{*}_{\Bar{E}}(D,A)$$
We can have multiple attention functions that satisfy this requirement, I will show one way we can construct a revealed attention function similar to how its done in the proof of Theorem 1.\\\\
Let $y$ be the $\succ^{*}$-maximal element in $A\cap \Bar{E}$. Then any alternative $x\in A$ such that $y\succ^{*}x$ will have zero probability because of C-WARP. The set of possible sets $D$ that satisfy the requirement, $\{D:[x\cup(A\cap \Bar{E})]\subseteq D\subseteq D_{x}\cap A\}$ is empty, so any attention function $\mu^{*}_{\Bar{E}}$ will satisfy the above equality.\\\\
For an arbitrary element $x\in A$ which is $\succ^{*}$-better than $y$, EDA and C-WARP say that they will have positive probability in the menu. Similar to the construction by \cite{Kovach23} we can ensure the existence of one such attention function by the following procedure: Let $n(x,A):=|\{D:[x\cup(A\cap \Bar{E})]\subseteq D\subseteq D_{x}\cap A\}|$. Then we can specify $\mu^{*}_{\Bar{E}}(D,A)=\frac{p(x,A)}{n(x,A)}$ for all $D\in \{D:[x\cup(A\cap \Bar{E})]\subseteq D\subseteq D_{x}\cap A\}$. This gives us the required functional form. The set $\Bar{E}$ is the revealed set of reference points which is unique. $\succ^{*}$ is the identified preferences unique up to the strict partial order $P$. $P$ uniquely gives us the preference of each alternative with respect to each reference points and the transitive closures that it implies.
\rule{2mm}{2mm}\medskip \newline\\\\
\textbf{Proof of Theorem 3}\\\\
\textit{Necessity:} \\\\
R-ASYM: If there exist $A\in\mathcal{X}\setminus\phi$ and $x,y\in A$ with $p(x,A)>0$ and $p(y,A)>0$, then either none of $x$ and $y$ are references, in which case the axiom is true because of the functional form, as proved in \cite{MM14}, or exactly one of them is a reference, and the other is preferred to it. In the latter case, the antecedent is true for the reference, since the probability of choice of the less preferred alternative (the reference) will increase if the other is removed, but removing the reference will not affect the choice probability of the better alternative.\\\\
R-IND: If there exists $A,B\in\mathcal{X}\setminus\phi$, $a,b\in A\cap B$ with $p(a,A)>0$ and $p(a,B)>0$, the functional form is such that the R-IND will hold. Same is true if instead of $a$ we consider the default $a^{*}$.
The proof is identical to the proof of the necessity of i-independence in \cite{MM14}.\\\\
NT: This axiom is trivially satisfied given the functional form of the RAM-UR-IRA.\\\\
EDA*: If $p(a^{*},A)=0$, for some $A\in\mathcal{X}\setminus\phi$ then there exists at least one alternative $x\in A$ such that $\gamma(x)=1$. Then $p(a^{*},\{x\})=0$ follows.\\\\
REG: Consider arbitrary $A\subseteq B\subseteq X$, then the functional form for any $x\in A$ gives us that $p(x,A)=\gamma(x)\prod_{y\in A; y\succ x}(1-\gamma(y))$. Similarly we get $p(x,B)=\gamma(x)\prod_{y\in B; y\succ x}(1-\gamma(y))$ for the set $B$. We can rewrite this as: $p(x,B)=\gamma(x)\prod_{y\in A; y\succ x}(1-\gamma(y))\prod_{y\in B\setminus A; y\succ x}(1-\gamma(y))$. Since all the attention parameters lie between zero and one, it is clear that $p(x,B)\leq p(x,A)$. So, REG is fulfilled.
\\\\
\textit{Sufficiency:}\\\\
I take the set 
$\Bar{E}=\{x\in X: p(x,A)=1\hspace{0.5mm}\text{for some}\hspace{0.5mm}A\in\mathcal{X}\setminus\phi\}$ to be the revealed set of reference alternatives.\\\\
Again, this set could be empty. If this set is empty, then EDA* and NT ensure that $p(a,A)>0$ for all $A\in\mathcal{X}\setminus\phi$ and $a\in A^{*}$. Now the R-ASYM and the R-IND will be equivalent to the i-ASYM and i-IND by \cite{MM14}, which gives us the required equivalence with the IRA model\footnote{An IRA model is a RAM-UR-IRA model with $E=\phi$.}. For the remainder of the proof, I assume non-emptyness of $\Bar{E}$.\\\\
I will use the notation IRA to denote any reference to \cite{MM14} in this proof. We define two more properties RIIA and EXP, which the above axioms together imply:\\\\
\textbf{Reference-Independence of Irrelevant Alternatives (RIIA):} For all $A\in\mathcal{X}\setminus\phi$ and $a\in A$: $p(a,A)=0$ implies $p(b,A\setminus\{a\})=p(b,A)$ for all $b\in A\setminus\{a\}$.\\\\
\textbf{Claim 1:} \textit{If a stochastic choice function $p:X\times\mathcal{X}\setminus\phi\rightarrow[0,1]$ satisfies REG, then it satisfies RIIA.}\\\\
\textbf{Proof:} By REG, for any alternative $c\in A^{*}\setminus\{a\}$ such that $p(c,A)>0$, we have $p(c,A)\leq p(c,A\setminus\{a\})$ and since summing over $p(c,A)$ for all $c\in A^{*}$ gives us one, they must be equal.
\\\\
\textbf{Expansion (EXP):} For all $A,B\in\mathcal{X}\setminus\phi$, $x_{1}\in A, x_{2}\in B$: $p(x_1,A)=p(x_2,B)=1$ implies either $p(x_1,A\cup B)=1$ or $p(x_2,A\cup B)=1$.
\\\\
\textbf{Claim 2:} \textit{If a stochastic choice function $p:X\times\mathcal{X}\setminus\phi\rightarrow[0,1]$ satisfies R-ASYM, R-IND and REG, then it satisfies EXP.}\\\\
\textbf{Proof:} By REG, $p(y, A\cup B)=0$ for all $y\in A\cup B)^{*}\setminus\{x_{1},x_{2}\}$. Suppose $p(x_1,A\cup B)>0$ and $p(x_2,A\cup B)>0$. Then by R-IND, we have $\frac{p(x_1,A\cup B\setminus\{x_2\})}{p(x_1,A\cup B)}=\frac{p(x_1,\{x_1\})}{p(x_1,\{x_1,x_2\})}$ and similarly for $x_2$.
But $p(x_1,\{x_1\})=p(x_2,\{x_2\})=1$ by REG. So, $\frac{p(x_1,A\cup B\setminus\{x_2\})}{p(x_1,A\cup B)}>1$ and $\frac{p(x_2,A\cup B\setminus\{x_1\})}{p(x_2,A\cup B)}>1$ which contradicts R-ASYM.\\\\
I proceed with the proof of Theorem 2 and start with the following two lemmas.

\begin{customlemma}{2}
$A\cap \Bar{E}=\phi$ for some $A\in\mathcal{X}\setminus\phi$ implies $p(a,A)>0$ for all $a\in A$.    
\end{customlemma}
\begin{proof} Suppose not. Then there exist a menu $A\in\mathcal{X}$ with $A\cap \Bar{E}=\phi$ and $a\in A$ such that $p(a,A)=0$.\\\\
By NT and RIIA there exists $x\in A$ such that $p(x,A)>0$.\\\\
By RIIA, $\frac{p(x,A\setminus\{a\})}{p(x,A)}=1$.\\\\
Repeatedly applying RIIA we can remove irrelevant alternatives\footnote{An alternative $a\in A\in\mathcal{X}\setminus\phi$ is irrelevant with respect to $A$ if $p(a,A)=0$.} from A to reach the menu $A^{'}=\{x:x\in A, p(x,A)>0\}$. Note that $p(x,A)=p(x,A^{'})$ for all $x\in A^{'}$.\\\\
Also note that since $A'\subseteq A$, $A'\cap\Bar{E} =\phi$. EDA* implies then $p(a^{*},A')>0$.
Due to the summation of choice probabilities, we have $p(a^*,A')=p(a^*,A)$. Since $a\in A$ is such that $p(a,A)=0$ we have $p(a^*,A)=p(a^*,A\setminus a)$\\\\
Using R-IND and NT, we have $1=\frac{p(a^*,A\setminus\{a\})}{p(a^*,A)}=\frac{p(a^*,\phi)}{p(a^*,a)}>1$\\\\
which is a contradiction.
\end{proof}\\\\
Denote by $\mathcal{X}^{'}$ the set of menus which contain only the elements not in $\Bar{E}$. I will first look only at the menus in the subsystem $\mathcal{X}^{'}$. Lemma 2 indicates that for any menu $A\in\mathcal{X}^{'}$, $p(a,A)>0$ for all $a\in A^{*}$.\\\\
Moreover, R-ASYM and R-IND reduce down to the i-asymmetry and the i-independence condition in the characterization of IRA, so in the subsystem $\mathcal{X}^{'}$, the required functional form equivalence is achieved due to the characterization result by \cite{MM14}. Also, the preferences are elicited using the i-regularity property which is also satisfied over this subsystem. Let the preferences elicited in such a way over alternatives in $X^{'}$ be $\succ^{'}$. Obviously, over this set the preference is a strict linear order.
\begin{customlemma}{3}
$A\cap \Bar{E}\neq\phi$ for some $A\in\mathcal{X}\setminus\phi$ implies $p(a^{*},A)=0$.    
\end{customlemma}
\begin{proof} 
Suppose not. Then there exists $A\in\mathcal{X}\setminus\phi$ such that $A\cap \Bar{E}\neq\phi$ and $p(a^{*},A)>0$. For $x\in A\cap \Bar{E}$, we have by REG that $p(a^{*},\{x\})>0$ or $p(x,\{x\})\neq 1$. Since $x\in \Bar{E}$, $p(x,B)=1$ for some $B\in\mathcal{X}\setminus\phi$. Applying REG leads to $p(x,\{x\})=1$ which is a contradiction.    
\end{proof}\\\\
I will now construct a binary relation over alternatives in $X$ and combine it with the preference $\succ^{'}$ generated over $X^{'}$ to form a preference over all alternatives. I will then show that this combination is a strict total order over all alternatives in $X$. For this, let us define the binary relation $\succ^{''}$ over $X$ as follows:\\\\
For all $a\in \Bar{E}$ and $b\in X$: $p(a,\{a,b\})=1$ implies $a\succ^{''}b$, and $p(a,\{a,b\})\neq 1$ implies $b\succ^{''} a$.\\\\
I can put these two binary relations together to get another binary relation: $\succ=\succ^{'}\cup\succ^{''}$.\\\\
\textbf{Claim 3: }\textit{The binary relation $\succ$ is a strict total order over $X$.}\\\\
\begin{proof}
Obviously, $\succ$ is complete by definition. Among the alternatives in $X^{'}$ it is clearly asymmetrical as well. We need to show asymmetry among alternatives where at least one of the alternatives is in $\Bar{E}$.\\\\
Consider an arbitrary $x\in\Bar{E}$ and arbitrary $y\in\Bar{E}$ preferences are only defined through $\succ^{''}$ and is asymmetric by definition.\\\\
Suppose $a\in \Bar{E}$ and $b\in \Bar{E}$. By EXP we have either $p(a,\{a,b\})=1$ or $p(b,\{a,b\})=1$. Without loss of generality, if $p(a,\{a,b\})= 1$, then $a\succ b$ but $b\nsucc a$. So, $\succ$ is asymmetric.\\\\
Now we show that $\succ$ is transitive as well. For this let us look at the following cases.\\\\
\textbf{Case 1:} $a\notin \Bar{E}$, $b\notin \Bar{E}$, $c\notin \Bar{E}$, $a\succ b$ and $b\succ c$. \\\\
The binary relation $\succ$ here is defined entirely by $\succ^{'}$ which is transitive as shown in the characterization of IRA, so transitivity of $\succ$ follows.\\\\
\textbf{Case 2:} $a\in \Bar{E}$, $b\in \Bar{E}$, $c\in X$, $a\succ b$ and $b\succ c$.\\\\
We have $p(a,\{a,b\})=1$ and $p(b,\{b,c\})=1$. By EXP, either $p(a,\{a,b,c\})=1$ or $p(b,\{a,b,c\})=1$ holds. \\\\
Suppose, $p(b,\{a,b,c\})=1$. Then by REG, $p(b,\{a,b\})=1$ follows. However, this contradicts the fact that $a\succ b$ and thus, $p(a,\{a,b,c\})=1$ should hold.\\\\
By REG again, we have $p(a,\{a,c\})=1$ and thus, $a\succ c$.\\\\
\textbf{Case 3:} $a\in \Bar{E}$, $b\notin \Bar{E}$ and $c\in \Bar{E}$, $a\succ b$ and $b\succ c$.\\\\
We have $p(a,\{a,b\})=1$ and $p(b,\{b,c\})>0$ (otherwise, $p(a^{*},\{b,c\})=1$ in contradiction to Lemma 3).\\\\
Since $c\in \Bar{E}$, we have by REG that $p(c,\{c\})=1$ holds. Using EXP, we have either $p(a,\{a,b,c\})=1$ or $p(c,\{a,b,c\})=1$.\\\\
If $p(c,\{a,b,c\})=1$, then by REG we have $p(c,\{b,c\})=1$ which is not true by assumption. So, $p(a,\{a,b,c\})=1$ should be the case.\\\\
By REG, we have $p(a,\{a,c\})=1$ and thus, $a\succ c$ follows.\\\\
\textbf{Case 4:} $a\in \Bar{E}$, $b\notin \Bar{E}$ and $c\notin \Bar{E}$, $a\succ b$ and $b\succ c$.\\\\
We have $p(a,\{a,b\})=1$ and $\frac{p(b,\{b\})}{p(b,\{b,c\})}=1$ (which holds due to $i-asymmetry^{*}$\footnote{This condition is defined by \cite{MM14} which says that the implication in \textit{i-asymmetry} goes both ways. They show that it is implied by their axioms.} which holds over the menus $\{b\}$ and $\{b,c\}$ since they are part of the subsystem $\mathcal{X}^{'}$ where all the IRA results apply). \\\\
\textbf{Claim:} $p(b,\{a,b,c\})=0$.\\\\
\begin{proof}
If not, then $p(b,\{a,b,c\})>0$. By REG, $p(b,\{a,b\})>0$ which is a contradiction.
\end{proof} \\\\
By RIIA and the above claim $p(c,\{a,c\})=p(c,\{a,b,c\})$. Suppose $p(c,\{a,b,c\})>0$, then by R-IND, $1=\frac{p(c,\{a,c\})}{p(c,\{a,b,c\})}=\frac{p(c,\{c\})}{p(c,\{b,c\})}$ but $\frac{p(c,\{c\})}{p(c,\{b,c\})}\neq1$ because of R-ASYM. Thus, it must be that $p(c,\{a,b,c\})=0$.\\\\
By RIIA, we have $p(c,\{a,c\})=p(c,\{a,b,c\})=0$. Applying Lemma 3, we get $p(a,\{a,c\})=1$. Thus, $a\succ c$ holds.\\\\
\textbf{Case 5:} $a\notin \Bar{E}$, $b\notin \Bar{E}$ and $c\in \Bar{E}$, $a\succ b$ and $b\succ c$.\\\\
We have $\frac{p(b,\{b\})}{p(b,\{a,b\})}>1$ (by i-regularity of the IRA result on the subsystem $\mathcal{X}^{'}$) and $p(b,\{b,c\})>0$ (by Lemma 3).\\\\
\textbf{Claim:} $p(a,\{a,b,c\})>0$.\\\\
\begin{proof}
Suppose not. Then $p(a,\{a,b,c\})=0$. By RIIA, $p(b,\{b,c\})=p(b,\{a,b,c\})>0$. By R-IND, $1=\frac{p(b,\{b,c\})}{p(b,\{a,b,c\})}=\frac{p(b,\{b\})}{p(b,\{a,b\})}>1$ which is not possible. So, $p(a,\{a,b,c\})>0$ follows.
\end{proof}\\\\
By R-IND, we have $\frac{p(a,\{a,c\})}{p(a,\{a,b,c\})}=\frac{p(a,\{a\})}{p(a,\{a,b\})}$. We can use R-IND since by Lemma 1 we have $p(a,\{a,b\})>0$ as well. Note that by NT, we have $p(a,\{a\})>0$ and thus, we must have $p(a,\{a,c\})>0$. Hence, $a\succ c$ follows.\\\\
\textbf{Case 6:} $a\notin \Bar{E}$, $b\in \Bar{E}$ and $c\in \Bar{E}$, $a\succ b$ and $b\succ c$.\\\\
We have $p(a,\{a,b\})>0$ (by Lemma 3) and $p(b,\{b,c\})=1$.\\\\
\textbf{Claim:} $p(a,\{a,b,c\})>0$.\\\\
\begin{proof}
Suppose not. then $p(a,\{a,b,c\})=0$. By RIIA we have $p(c,\{a,b,c\})=p(c,\{b,c\})=0$ and $p(b,\{a,b,c\})=p(b,\{b,c\})=1$. Applying RIIA again gives us $p(b,\{a,b\})=p(b,\{a,b,c\})=1$ which is not possible, so $p(a,\{a,b,c\})>0$.
\end{proof}\\\\
By R-IND, $\frac{p(a,\{a,c\})}{p(a,\{a,b,c\})}=\frac{p(a,\{a\})}{p(a,\{a,b\})}$. This implies $p(a,\{a,c\})>0$ because $p(a,\{a\})>0$ and $p(a,\{a,b\})>0$. Thus, $a\succ c$ holds.\\\\
\textbf{Case 7:} $a\notin \Bar{E}$, $b\in \Bar{E}$ and $c\notin \Bar{E}$, $a\succ b$ and $b\succ c$.\\\\
We have $p(a,\{a,b\})>0$ (by Lemma 3) and $p(b,\{b,c\})=1$.\\\\
\textbf{Claim:} $p(a,\{a,b,c\})>0$.\\\\
\begin{proof}
Suppose not. then $p(a,\{a,b,c\})=0$. By RIIA we have $p(c,\{a,b,c\})=p(c,\{b,c\})=0$ and $p(b,\{a,b,c\})=p(b,\{b,c\})=1$. Applying RIIA again gives us $p(b,\{a,b\})=p(b,\{a,b,c\})=1$ which is not possible, so $p(a,\{a,b,c\})>0$.
\end{proof}\\\\
Note that $p(c,\{a,b,c\})=0$ by REG since $p(c,\{b,c\}=0$ by assumption.\\\\
Since $a\notin \Bar{E}$, $p(a,\{a,b,c\})<1$. Since $p(c,\{a,b,c\})=0$ and $p(a^{*},\{a,b,c\})=0$ (by Lemma 3), we have $p(b,\{a,b,c\})>0$.\\\\
It follows from $\frac{p(b,\{b,c\})}{p(b,\{a,b,c\}}>1$ and R-ASYM that $\frac{p(a,\{a,c\})}{p(a,\{a,b,c\})}=1$ holds. By R-IND, we have $\frac{p(a,\{a\})}{p(a,\{a,c\})}=\frac{p(a,\{a,c\})}{p(a,\{a,b,c\})}=1$. In the subsystem containing the menu $\{a,c\}$, $i-asymmetry^{*}$ holds and thus, $\frac{p(c,\{c\})}{p(c,\{a,c\})}>1$ leads to $a\succ c$.\\\\
In view of the above cases we have that $\succ$ is transitive and thus, a strict total order over $X$.
\end{proof}
\\\\
We now define $\gamma:X\rightarrow(0,1]$ by $\gamma(a)=p(a,\{a\})$ for each $a\in X$. Note that $\gamma(a)=1$ for all $a\in \Bar{E}$ and $\gamma(b)\in(0,1)$ for all $b\notin \Bar{E}$. $\Bar{E}$ will be identified here as the set $E$ of the RAM-UR-IRA.\\\\
%
%
\textbf{Claim 4:} For all $A\in \mathcal{X}\setminus\phi$ and $a,b\in A$: $p(a,A)>0$, $p(b,\{b\})\neq 1$ and $p(a,A\setminus\{b\})>p(a,A)$ imply $\frac{p(a,A\setminus\{b\})}{p(a,A)}=\frac{1}{1-p(b,\{b\})}$ for all $a\in A^{*}$.
\\\\
\begin{proof}
This claim is similar to the one in the proof of the first theorem in \cite{MM14}. Proving this claim will allow us to recover the functional form of the IRA structure.\\\\
For $a^{*}$, the required equality translates to $\frac{p(a^{*},A\setminus\{b\})}{p(a^{*},A)}=\frac{1}{1-p(b,\{b\})}$. Note that $\frac{1}{1-p(b,\{b\})}=\frac{p(a^{*},\phi)}{p(a^{*},\{b\})}$.
which is true by the implementation of R-IND.\\\\
For $a,b\in A$ such that $\frac{p(a,A\setminus\{b\})}{p(a,A)}>1$. Here, RIIA implies that $p(b,A)>0$.\\\\
We can apply R-ASYM, to get $p(b,A\setminus\{a\})=p(b,A)$\\\\
By REG we get $p(a,\{a,b\})>0$.\\\\
By R-IND we have, 
$$\frac{p(a,A\setminus\{b\})}{p(a,A)}=\frac{p(a,\{a\})}{p(a,\{a,b\})}=\frac{1-p(a^{*},\{a\})}{1-p(b,\{a,b\})-p(a^{*},\{a,b\})}$$.
Applying R-ASYM,
$$\frac{p(a,A\setminus\{b\})}{p(a,A)}=\frac{1-p(a^{*},\{a\})}{1-p(b,\{b\})-p(a^{*},\{a,b\})}$$
If $p(a,\{a\})=1$, then we get the required equation by REG. If not we have $a\notin \Bar{E}$ so by EDA* we have $p(a^{*},\{a,b\})>0$. We can then use the following equation from the first part of this proof:
$$p(a^{*},\{a\})=\frac{p(a^{*},\{a,b\})}{1-p(b,\{b\})}$$
Putting this above we will get the required equality:
$$\frac{p(a,A\setminus\{b\})}{p(a,A)}=\frac{1}{1-p(b,\{b\})}$$
\end{proof}\\\\
Now I will look at different set of menus and alternatives to show that we get the required functional form. \\\\
For $a\in \Bar{E}$, define $A=\{a\}\cup\{x:x\in X, a\succ X\}$. By definition of $\succ$, $p(a,\{a,b\})=1$ for all $b\in A$. So, by REG, $p(a,A)=1=\delta(a)$. As required.\\\\
Let $A\in\mathcal{X}$ be a menu such that there exists atleast one $x\in A\cap \Bar{E}$. Let $a\in A\cap \Bar{E}$ such that $a\succ b$ for any $b\in A\cap \Bar{E}\setminus\{a\}$, essentially, $a$ is the most preferred alternative in $A\cap \Bar{E}$. \\\\
For any alternative $b\in A$ such that $a\succ b$, $p(b,A)=0$ by REG. Also by REG, $p(a^*,A)=0$ as required.\\\\
\textbf{Claim:} In the above case, $p(a,A)>0$.\\\\
\begin{proof}
Suppose not. Then $p(a,A)=0$.\\\\
Since we have assumed $p(a,A)=0$, $p(c,A)=p(c,A\setminus\{a\})$ follows for all $c\in A\setminus\{a\}$. We can iteratively remove all alternatives in $A$ that are worse than $a$ and $a$ itself to reach $A^{'}=\{b\in A:b\succ a\}$.\\\\
Since, $a$ is the most preferred alternative in $\Bar{E}$ that is in $A$, $A^{'}\cap \Bar{E}=\phi.$. So by EDA*, $p(a^{*},A^{'})>0$. \\\\
Since $p(b,A)=p(b,A^{'})$ for all $b\in A^{'}$ and $p(c,A)=0$ for all $c\in A\setminus (A^{'})$, $p(a^{*},A)>0$ follows in contradiction to REG.
\end{proof}\\\\
\textbf{Claim:} If $c\succ a$ and $c\in A$, then $p(c,A)>0$.\\\\
\begin{proof}
Suppose, there exists $c\in A$ such that $p(c,A)=0$ and $c\succ a$. Since $p(a,A)>0$, we have by R-IND $1<\frac{p(a,\{a\})}{p(a,\{a,c\})}=\frac{p(a,A\setminus\{c\})}{p(a,A)}$ which is incompatible with RIIA. So, it must be that $p(c,A)>0$ for all $c\succ a$.
\end{proof}\\\\
I can now iteratively apply Claim 4 similar to how it was done in \cite{MM14} to get the required equivalence with the functional form. We have already considered menus without any alternatives in $\Bar{E}$ through the subsystem $\mathcal{X}^{'}$. The rest of the cases have been considered above.
$E$ is identified uniquely as the $\Bar{E}$ and preferences as specified above are identified uniquely by $\succ$.
\rule{2mm}{2mm}\medskip \newline
\subsection*{Appendix 2}
In this section, I discuss the Random Utility Model representation of the RAM-UR-IRA and analyze how the representation changes as the set of reference alternatives $E$ changes. \\\\
Denote by $\mathcal{L}(X^{*})$ the set of all strict total orders over $X^{*}$ with typical element $R$. A probability distribution over $\mathcal{L}(X^{*})$ is denoted by $\nu$.
\begin{definition}
A stochastic choice function $p:X^{*}\times \mathcal{X}\setminus\phi$ has a \textbf{Random Utility Model (RUM)} representation if there exists a distribution $\nu$ over the set of all strict total orders on $X^{*}$, $\mathcal{L}(X^{*})$ such that:
$$p(x,A)=\sum_{R\in \mathcal{L}(X^{*}): xR y\forall y\in A^{*}\setminus\{x\}}\nu(R)$$
for all $x\in A^{*}$.
\end{definition}
The representation is very similar in construction to the one provided by \cite{MM14} for their IRA model which is a RUM too. The construction for the IRA goes as follows:
\begin{itemize}
    \item For any alternative $a\in X$, set $\nu(\{R:aR a^{*}\})=\gamma(a)$. 
    \item If $a\succ b$, then for an $R$ such that $bR a$, $aR a^{*}$ and $bR a^{*}$, set $\nu(R)=0$.
    \item For any two alternatives $a$ and $b$, $\nu(\{R:aRa^{*}\hspace{1mm}\text{and}\hspace{1mm}bRa^{*}\})=\gamma(a)\gamma(b)$.
\end{itemize}
The above restrictions continue to hold for the RAM-UR-IRA but we will also need more restrictions for our model:
\begin{enumerate}
    \item A reference is always considered if available. So, we have $\nu(\{R:a^{*}Ra\})=0$ for all $a\in E$.
    \item If $a\in E$ and $a\succ b$ for $b\in X$, set $\nu(\{R:b\succ a\})=0$.
\end{enumerate}
(1) here is implied by the first point in the construction by \cite{MM14}. It is reiterated to emphasize that all strict orders where the default beats a reference alternative are essentially removed from the support of the RUM.\\\\
Note that as we add more alternatives to the set $E$, more and more strict total orders are removed from the support. As $E$ becomes the set $X$, only the strict linear order, which is the preference of the agent is in the support, and the agent is rational. On the other hand, if $E$ is the empty set, the first set of restrictions will be the only ones that apply, and we are back to the \cite{MM14} model. Looking at the RUM representation makes the nestedness properties of the RAM-UR-IRA clearer.\\\\
Note finally that the GRAM-UR or the RAM-UR might not be RUM because they fail to satisfy regularity which is a necessary condition for a stochastic choice function to be a RUM, (\cite{Falmagne}). This is the case because of the lack of structure on the attention function. The lack of structure of the general models allows us to construct attention functions that gives us the means to explain different behavioral observations. One such empirically observed behavioral pattern that it allows us to explain is that the status quo (references) is more likely to be chosen in larger sets (Choice Overload) similar to how \cite{Kovach23} showed in their model using the Luce Random Attention structure. 


\newpage
\begin{thebibliography}{99}
\bibitem[Bossert and Sprumont(2009)]{Bossert09}
Bossert, W. and Y. Sprumont (2009): ``Non-Deteriorating Choice," \textit{Economica}, 76(302), 337–363. 
%
\bibitem[Brady and Rehbeck(2016)]{Brady}
Brady, R. L., and J. Rehbeck (2016): “Menu-Dependent Stochastic Feasibility,” \textit{Econometrica}, 84(3), 1203–1223. 
%
\bibitem[Cattaneo et al.(2020)]{Cattaneo}
Cattaneo, M. D., Ma, X., Masatlioglu, Y. and E. Suleymanov (2020): ``A Random Attention Model," \textit{Journal of Political Economy}, 128(7), 2796-2836.
%
\bibitem[Dawes and Brown(2005)]{Dawes}
Dawes, P. L., and J. Brown (2005): ``The Composition of Consideration and Choice Sets in Undergraduate University Choice: An Exploratory Study," \textit{Journal of Marketing for Higher Education}, 14(2), 37–59.
%
\bibitem[Dean et al.(2017)]{Dean17}
Dean, M., Kıbrıs, Ö. and Y. Masatlioglu (2017): ``Limited attention and status quo bias," \textit{Journal of Economic Theory}, 169, 93-127.
%
\bibitem[Dew and Kwon(2009)]{Dew}
Dew, L., and W-S. Kwon (2009): ``Exploration of Apparel Brand Knowledge: Brand Awareness, Brand Association, and Brand Category Structure," \textit{Clothing and Textiles Research Journal}, 28(1), 3-18.
%
\bibitem[Dhar(1997)]{Dhar}
Dhar R. (1997): ``Consumer Preference for a No-Choice Option," \textit{Journal of Consumer Research}, 24(2), 215–231.
%
\bibitem[Falmagne(1978)]{Falmagne}
Falmagne, J. C. (1978): ``A representation theorem for finite random scale systems," \textit{Journal of Mathematical Psychology}, 18(1), 52–72.

%
\bibitem[Horan(2019)]{Horan}
Horan, S. (2019): ``Random Consideration and Choice: A Case Study of “Default” Options," \textit{Mathematical Social Sciences}, 102, 73-84.
%
\bibitem[Huber et al.(1982)]{Huber}
Huber, J., Payne, J. W., and C. Puto (1982): ``Adding Asymmetrically Dominated Alternatives: Violations of Regularity and the Similarity Hypothesis," \textit{Journal of Consumer Research}, 9(1), 90–98.
%
\bibitem[Kovach and Suleymanov(2023)]{Kovach23}
Kovach, M. and E. Suleymanov (2023): ``Reference Dependence and Random Attention," \textit{Journal of Economic Behavior and Organization}, 215, 421-441.

%
\bibitem[Lleras et al.(2017)]{Lleras}
Lleras, J. S., Masatlioglu, Y., Nakajima, D. and E. Y. Ozbay (2017): ``When More is Less: Limited Consideration," \textit{Journal of Economic Theory}, 170, 70-85.
%
\bibitem[Manzini and Mariotti(2014)]{MM14}
Manzini, P. and M. Mariotti (2014): ``Stochastic choice and consideration sets,” \textit{Econometrica}, 82, 1153–1176.    
%
\bibitem[Masatlioglu and Ok(2005)]{MO05}
Masatlioglu, Y. and E. A. Ok (2005): ``Rational choice with status quo bias," \textit{Journal of Economic Theory}, 121(1), 1-29.
%
\bibitem[Masatlioglu et al.(2012)]{Masatlioglu12}
Masatlioglu, Y., Nakajima, D. and E. Y. Ozbay (2012): ``Revealed Attention," \textit{American Economic Review}, 102 (5): 2183-2205.
%
\bibitem[Manzini and Mariotti(2007)]{MM07}
Manzini, P. and M. Mariotti (2007): ``Sequentially rationalizable choice,” \textit{American
Economic Review}, 97, 1824–1839.
%
\bibitem[Masatlioglu and Ok(2014)]{MO14}
Masatlioglu, Y. and E. A. Ok (2014): ``A Canonical Model of Choice with Initial Endowments," \textit{The Review of Economic Studies}, 81(2 (287)), 851–883.
%
\bibitem[Nedungadi(1990)]{Nedungadi}
Nedungadi, P. (1990): ``Recall and Consumer Consideration Sets: Influencing Choice without Altering Brand Evaluations," \textit{Journal of Consumer Research}, 17(3), 263–276. 
%
\bibitem[Ok et al.(2015)]{OK15}
%
Ok, E. A., Ortoleva, P. and G. Riella (2015): ``Revealed (P)Reference Theory." \textit{American Economic Review}, 105(1): 299-321.
%
\bibitem[Rosen et al.(2008)]{Rosen}
Rosen, D. E., Curran, J. M., and T. B. Greenlee (1998): ``College Choice in a Brand Elimination Framework: The High School Student’s Perspective," \textit{Journal of Marketing for Higher Education}, 8(3), 73–92. 
%
\bibitem[Sagi(2006)]{Sagi06}
Sagi, J. S. (2006): ``Anchored preference relations," \textit{Journal of Economic Theory}, 130(1), 283-295.
%
\bibitem[Simon's(1956)]{Simon}
Simon, H.A. (1956): ``Rational choice and the structure of the environment," \textit{Psychological Review}, vol. 63, March.
%
\bibitem[Szpilrajn(1930)]{Szpilrajn}
Szpilrajn, E. (1930): ``Sur l'extension de l'ordre partiel," \textit{Fund. Math.}, (16), 386-389.
%
\bibitem[Tapkı(2007)]{Tapki07}
Tapkı, İ. G. (2007): ``Revealed incomplete preferences under status-quo bias," \textit{Mathematical Social Sciences}, 53(3), 274-283.
%
\bibitem[Tserenjigmid(2019)]{Tserenjigmid19}
Tserenjigmid, G. (2019): ``Choosing with the worst in mind: A reference-dependent model," \textit{Journal of Economic Behavior \& Organization}, 157, 631-652. 
%
\bibitem[Wilson(2008)]{Wilson}
Wilson, C.J. (2008): ``Consideration Sets and Political Choices: A Heterogeneous Model of Vote Choice and Sub-national Party Strength," \textit{Political Behavior}, 30, 161–183.
\end{thebibliography}
\end{document}